\documentclass[a4paper,11pt]{article}
\usepackage{pos}
\usepackage[utf8]{inputenc}

\usepackage{amsmath}
\usepackage{amsfonts}
\usepackage{bbm}
\usepackage{color}
\usepackage{comment}
\usepackage{graphicx}
\usepackage{epstopdf}
\usepackage{tikz}
\usepackage{graphicx}

\usepackage[english]{babel}

\makeatletter

\newcommand{\be}{\begin{equation}}
\newcommand{\ee}{\end{equation}}
\newcommand{\beq}{\begin{eqnarray}}
\newcommand{\eeq}{\end{eqnarray}}
\newcommand{\ba}{\begin{align}}
\newcommand{\ea}{\end{align}}

\newcommand{\mathi}{{\rm i}}

\title{If the universe were curved and noncommutative}
\ShortTitle{}

\author*[a]{S.~A.~Franchino-Viñas}
\author[b,c]{S.~Mignemi}

\affiliation[a]{Helmholtz-Zentrum Dresden-Rossendorf,\\
Bautzner Landstraße 400, 01328 Dresden, Germany.}

\affiliation[b]{Dipartimento di Matematica e Informatica, Università di Cagliari,
\\
viale Merello 92, 09123 Cagliari, Italy.}

\affiliation[c]{
INFN, Sezione di Cagliari,
\\
Cittadella Universitaria, 09042 Monserrato, Italy.}

\emailAdd{s.franchino-vinas@hzdr.de}
\emailAdd{smignemi@unica.it}

\abstract{
We review recent discussions concerning the definition of a quantum field theory in
a curved and noncommutative space, the Snyder--de Sitter space.
For a quartic self-interacting scalar field in a spacetime of arbitrary dimension,
we show how to perturbatively define the classical action in the small-deformation regime
and give its explicit first-order expression.
Afterwards, we compute the divergences of the one-loop effective action
for the two- and four-dimensional cases.
These are employed to calculate the beta functions of the couplings
and to numerically analyze the corresponding renormalization group flow.
Depending on the initial conditions of the couplings,
in four dimensions it is possible to obtain an asymptotic-free theory
or a flip in the sign of the cosmological constant as a consequence of the running.

}

\FullConference{Corfu Summer Institute 2023 "School and Workshops on Elementary Particle Physics and Gravity" (CORFU2023)\\
 23 April - 6 May , and 27 August - 1 October, 2023\\
Corfu, Greece\\}

\begin{document}
\maketitle


\section{Introduction}

{Quantum gravity} consists of two words, a considerable amount of ideas and much fewer successes.
A tempting approach to quantum gravity is to interpret gravity as an effective quantum field theory (QFT),
where several of the thought-to-be inconveniences simply become inherent features of the theory~\cite{dePaulaNetto:2021axj, Donoghue:2022eay};
in this way, for example, one is able to gracefully compute (classical) scattering amplitudes between compact objects~\cite{Dlapa:2023hsl, Klemm:2024wtd}.
Other strategies in quantum gravity include the idea that it could display a fixed point in the sense of the renormalization group~\cite{Bonanno:2020bil, Fraaije:2022uhg, Gies:2022ikv},
that the Einstein--Hilbert action should be modified to incorporate higher-order terms~\cite{Buccio:2024hys}, that supersymmetry should play a key r\^ole (as in stringy supergravity theories~\cite{Montero:2022ghl}) and that the nature of spacetime is much intrincated, as in group field theory~\cite{Marchetti:2022nrf} or spin foam theories~\cite{Borissova:2022clg, Jercher:2023rno}.

An alternative, albeit related path is given by noncommutative QFTs~\cite{Snyder:1946qz}.
In this approach, the underlying spacetime is promoted to a nontrivial algebra,
whose commutation relations imply a granularity of spacetime at high energies and introduce several novelties at the level of the field theory~\cite{Douglas:2001ba, Szabo:2001kg}.
From a mathematical perspective, it has given rise to a new branch in mathematics~\cite{Connes:1994yd};
some late developments include the study of the quantum Poincaré symmetry of $\rho$-Minkowski~\cite{Fabiano:2023uhg}
and the application of spectral triples, both to gauge theories~\cite{Iseppi:2022nyy} and to the generation of skew torsion from a torsionless connection~\cite{Martinetti:2024umx}.

From a phenomenological point of view, noncommutative theories offer a fertile playground (see the reviews~\cite{Addazi:2021xuf, AlvesBatista:2023wqm} and references therein).
Indeed, the involved structures admit expansions in the so-called deformation parameters;
in their turn, these expansions account for new predictions, which can be readily tested with observations.
A physical prerequisite for this is that the theory should incorporate at least the basic features of our universe;
in particular, if one would desire to consider cosmological aspects, it seems unavoidable to accomodate the curvature displayed by our universe at large scales~\cite{planck18}.
This has inspired several approaches, among them the use of Poisson-Lie algebras~\cite{Ballesteros:2022nyx},
the implementation of a fuzzy de Sitter space~\cite{Brkic:2024sud},
the use of embeddings in higher dimensional spaces as a way to construct de Sitter structures~\cite{Frob:2020ctp},
and generalizations of the Snyder spaces including curvature~\cite{KowalskiGlikman:2004kp, Mignemi:2008fj, Mignemi:2009zz}.

In this manuscript, the geometric structure will be determined by a variant of the latter, the Snyder--de Sitter (SdS) space.
As we will see, in this space one can conserve an underformed action of the Lorentz generators,
thanks to which the contact with standard QFTs is made more straightforward than in other theories.
We will closely follow the results obtained in Refs.~\cite{Franchino-Vinas:2019nqy, Franchino-Vinas:2019lyi, Franchino-Vinas:2021bcl} for a quartic self-interacting, real scalar QFT
defined in Euclidean SdS,
which in their turn relied on the previous developments in flat Snyder spaces~\cite{ Meljanac:2017ikx, Franchino-Vinas:2018jcs}
and general features of QFT in curved space~\cite{Parker:2009}.
As we will see, a double expansion in the curvature and the noncommuative deformation is already sufficient to derive some novel predictions
and understand some peculiarities of these theories.


\section{Snyder--de Sitter space}
Let us start by explicitly writing down the algebras which define the Euclidean SdS algebra in arbitrary $D$ dimensions.
First of all, we would like to keep an undeformed (Euclidean) $\mathfrak{so}(D)$ algebra for the Lorentz generators $J_{ij}$:
\begin{align}\label{eq:lorentz_algebra}
 \begin{split}
[J_{ij}, J_{kl}]&=\mathi (\delta_{ik}J_{jl}-\delta_{il}J_{jk}-\delta_{jk}J_{il}+\delta_{jl}J_{ik}).
 \end{split}
 \end{align}
Inspired by the usual considerations, it is natural to regard these generators as a composition of the momentum and position operators;
bearing in mind that the commutation relations between the latter may turn out to be noncanonical,
we consider the Ansatz
\begin{align}\label{eq:lorentz_generator}
 J_{ij}\equiv \frac{1}{2}(\hat x_i\hat p_j-\hat x_j\hat p_i+\hat p_j\hat x_i-\hat p_i\hat x_j).
\end{align}
In these expressions, we use Latin indices $i=0,\cdots D$ because we are interested in discussing a $D$-dimensional Euclidean space;
in particular, $\hat x_0$ corresponds to the Euclidean time operator, while $\hat x_i$ with $i=1\cdots D$ are the corresponding spatial operators.

One can then verify that the Eqs.~\eqref{eq:lorentz_algebra} and~\eqref{eq:lorentz_generator}
are compatible with the following deformation of the phase space algebra into a quadratic algebra,
which includes two parameters, $\alpha$ and $\beta$ (see below for their interpretation):
\begin{align}\label{eq:phase_space}
\begin{split}
 &[\hat{x}_i,\hat{x}_j]={\rm i}\beta^2 J_{ij},\qquad [\hat{p}_i,\hat{p}_j]={\rm i}\alpha^2 J_{ij},\\
 &[\hat{x}_i,\hat{p}_j]={\rm i}[\delta_{ij}+\alpha^2 \hat{x}_i\hat{x}_j+\beta^2\hat{p_j}\hat{p}_i+\alpha\beta (\hat{x}_j\hat{p}_i+\hat{p_i}\hat{x}_j)].
 \end{split}
\end{align}
The only remaining commutators are those which fix the action of the Lorentz generators on the position and momentum operators,
which  correspond to the simple vectorial case,
\begin{align}\label{eq:lorentz_action}
 \begin{split}
 [J_{ij}, \hat{p}_{k}]&=i (\delta_{ik} \hat{p}_{j}-\delta_{kj} \hat{p}_{i}),\\
 [J_{ij}, \hat{x}_{k}]&=i (\delta_{ik} \hat{x}_{j}-\delta_{kj} \hat{x}_{i}).
 \end{split}
 \end{align}

The choice that we have made in Eq.~\eqref{eq:phase_space} encompasses several desired properties.
The most important one is probably the exchange symmetry between the dimensionless position and momentum operators:
\begin{align}\label{eq:born_duality}
\alpha\hat x_i \leftrightarrow\beta\hat p_i.
\end{align}
Indeed, this property clearly restricts the space of algebras that we could have considered;
for instance, it is satisfied by the canonical quantization in flat space
but will clearly not be preserved in a general curved (commutative) space.
This idea, proposed already by Born in 1949~\cite{Born:1949} and subsequently baptized as Born's reciprocity,
might seem rather far-fetched, given that it would prevent us from including, for example, position-dependent potentials in our theories;
however, it has revealed to be the departure point for some of the most significant advances
in foundational questions related to the construction of QFTs in noncommutative spaces~\cite{Langmann:2002cc, Grosse:2005da}.

Coming back to the meaning of the deformation parameters,
notice that for vanishing $\beta$ the phase space algebra~\eqref{eq:phase_space}
reduces to the one corresponding to de Sitter space (if $\alpha^2>0$),
where $\alpha$ plays the role of the inverse of the de Sitter radius $l$,
which can be readily linked to the cosmological constant $\Lambda$;
for arbitrary $D>2$, the relation reads $\Lambda=\tfrac{(D-1)(D-2) \alpha^2}{2 }$.
For this reason, we will associate $\alpha$ with the cosmological constant or,
more precisely, with a curvature in spacetime, even in the completely deformed setup with $\beta\neq0$.

On the other hand, for $\alpha=0$ the deformation corresponds to the original definition of the Snyder space~\cite{Snyder:1946qz}.
Curiously enough, there exists one family of noncanonical, linear maps between the SdS algebra and the Snyder space,
which are parameterized by $t\in\mathbb{R}$, if hermiticity is to be preserved:
\begin{align}\label{eq:2snyder}
 \hat{x}_i=:X_i+t\, \frac{\beta}{\alpha} P_i, &\qquad \hat{p}_i=:(1-t)P_i -\frac{\alpha}{\beta} X_i.
\end{align}
Indeed, a straigthforward replacement of Eq.~\eqref{eq:2snyder} in \eqref{eq:phase_space}
leads to the usual Snyder phase space in the capital variables $(X_i,P_j)$,
\begin{equation}\label{eq:Snyder}
 [X_i,X_j]={\rm i}\beta^2 J_{ij}, \qquad [P_i,P_j]=0,\qquad [X_i,P_j]={\rm i}(\delta_{ij}+\beta^2P_iP_j),
\end{equation}
regardless of the value of $t$.
At this point, for simplicity we will consider $t=0$; later on, we will argue that this is indeed the best choice one could make.
In this way, we can use the already known machinery of realizations in Snyder spaces, which are lifted to realizations in SdS through the map~\eqref{eq:2snyder}.
In particular, we will employ the following symmetrized (Hermitian) realization in terms of canonical operators\footnote{By canonical, we mean of course that they satisfy the algebra $[x_i,p_j]=\mathi \delta_{ij}$. Moreover,  we also assume the usual realization on Lebesgue square integrable functions, i.e. $p_j\equiv-\mathi\partial_j$ and $x_j\equiv x_j$.} $x_i$ and $p_j$,
\begin{align}\label{eq:2canonical}
 \begin{split}
P_i&\equiv p_i\equiv-{\rm i}\partial_i, \\
 X_i&\equiv x_i+\frac{\beta^2}{2} (x_jp^jp_i+p_ip^jx_j).
 \end{split}
\end{align}
A simplification entailed by this realization is that the corresponding Weyl--Moyal (also called star) product is known, see Ref.~\cite{Meljanac:2017ikx};
the basic star product ($\star$) of exponentials is given by
\begin{align}\label{eq:star_product}
 e^{{\rm i}k\cdot x}\star e^{{\rm i} q \cdot x}:=\frac{e^{{\rm i}D(k, q)\cdot x}}{(1-\beta^2 k\cdot q)^{(D+1)/2}},
\end{align}
where the explicit expression for the momentum composition $D_{j}$ is
\begin{align}
 D_{j}(k,q):=\frac{1}{1-\beta^2 k \cdot q} \left[\left(1- \frac{\beta^2 k \cdot q}{1+\sqrt{1+\beta^2 k^2}}\right)k_{j}+ \sqrt{1+\beta^2 k^2} q_{j}\right].
\end{align}
Using this Weyl--Moyal product, one can as usual reinterpret the deformation of the operators' algebra in terms of the deformed product $\star$.
Evidently, this product is noncommutative, so that it is consistent with the first expression in Eq.~\eqref{eq:Snyder};
more surprising is the fact that it is also nonassociative and one has to be careful with the gathering of the different elements in a given expression.
Notably, this star product is closed~\cite{Meljanac:2017ikx}, i.e.
\begin{align}\label{eq:star_integral}
 \int {\rm d}^Dx f(x) \star g(x) = \int {\rm d}^Dx f(x) g(x).
\end{align}

Unluckily, it's not all roses and sunshine.
The map~\eqref{eq:2snyder} is singular in both $\alpha$ and $\beta$,
so that we should be careful if we were try to conceptually build an effective field theory in these deformation parameters.
The best we can do is to fix $t\equiv 0$, for which the Snyder limit ($\beta\to0$) is expected to be smooth.\footnote{One could actually use  Born's reciprocity, which is broken by the map~\eqref{eq:2snyder}, to alternatively obtain a smooth de Sitter limit; we are not going to pursue this idea further in this review.}
As a matter of completion, notice that in this framework the original SdS operators can be realized as
\begin{align}\label{eq:symmetrization}
 \hat x_i&\equiv x_i+\frac{\beta^2}{2} (x_jp^jp_i+p_ip^jx_j),\\
 \hat{p}_i&\equiv p_i-\frac{\alpha}{\beta} x_i-\frac{\alpha \beta}{2} (x_jp^jp_i+p_ip^jx_j).
\end{align}

\section{Building a quantum field theory on Snyder--de Sitter}
\subsection{Action}
With the elements developed in the previous section it is not arduous to conceive a suitable action for a quantum field in SdS.
Following an orthodox path, a natural guess for a free scalar field is a ``$p^2+m^2$'' term;
as long as hermiticity (or positivity) is desired, it can be checked that one is left with just one option~\cite{Franchino-Vinas:2021bcl}:
\begin{align}\label{eq:free_action}
 S_K:=\frac{1}{2}\int {\rm d}^Dx \,\Big\lbrace \varphi\, \left[(\hat p (\hat p \,\varphi)\right]+ m^2 \varphi^2 \Big\rbrace \, .
\end{align}
Indeed, the remaining alternative associations in the first term lead to undesirable imaginary terms
which could provoke unacceptable instabilities at the level of the free theory.
In terms of the canonical operators and expanding to second order in the deformation parameters $\alpha$ and $\beta$,
the kinetic part of the action can be written as
\begin{align}\label{eq:kinetic}
 \begin{split}
S_K&\approx
 \frac{1}{2}\int {\rm d}^Dx \,\varphi\, \Bigg(p^2+\omega^2 x^2
 +\alpha^2_{\rm eff} x_jp^jp^ix_i+ m^2_{\text{eff}}\Bigg)\, \varphi,
 \end{split}
\end{align}
where, to simplify the notation in the following,
we have introduced labels for all the effective couplings that appear in the Lagrangian,
including an effective mass $m_{\rm eff}$, an effective curvature $\alpha_{\rm eff}^2$ and the oscillator frequency $\omega$:
\begin{align}
 \label{eq:meff}
 m^2_{\text{eff}}:&=m^2-\frac{\alpha^2}{2}D(D+1),
 \\
 \omega^2&=\frac{\alpha^2}{\beta^2},
 \\
 \alpha_{\rm eff}^2 :&= 2\alpha^2.
\end{align}
The negative contribution to the effective mass in Eq.~\eqref{eq:meff} is reminiscent of (but not completely equivalent to) the situation in ordinary de Sitter space, which can be understood in terms of representation theory~\cite{Marolf:2012kh}. Still, if for some reason the effective mass vanishes (or becomes negative), a perturbative expansion neglecting some terms in Eq.~\eqref{eq:kinetic} might be precluded, a situation once more similar to the de Sitter case~\cite{LopezNacir:2019ord}.

Before proceeding, let us make the following phenomenological observation:
as a typical mass for a scalar field, we could take the mass of the pion, which is approximately $139\, \text{MeV}$
(we could also take a typical GeVish baryonic mass without altering radically the following discussion).
On its turn, it should be $\alpha\sim 10^{-33}\, \text{eV}$,
if we were to associate it with the observed value of the cosmological constant (see the discussion before Eq.~\eqref{eq:2snyder} above),
while we could assume $\beta\sim 10^{-29}\,\text{eV}^{-1}$, i.e. of the order of the Planck length.
We thus observe a significant cancellation of scales, so that $\omega\sim 10^{-4}\,\text{eV}^2$.
This is the reason why we have retained the quotient $\alpha/\beta$ in expression~\eqref{eq:kinetic},
taking it to be of zeroth order in the deformation expansion.
Curiously enough, this quotient regulates the intensity of a harmonic term, which was a crucial ingredient in the vulcanization of the Grosse-Wulkenhaar model \cite{Grosse:2005da}.
In other words, since this term does not appear for the simple Snyder model,
it could be conjectured that adding curvature could be a way of regularizing a certain set of noncommutative theories, what was already noted in Ref.~\cite{Buric:2009ss}.

Coming back to the building of the action, the Weyl-Moyal product becomes extremely useful when we want to consider a self-interacting theory.
As a simple example, let us analyze a quartic self-coupling and take one particular ordering (recall that the product is nonassociative).
If we choose
\begin{align}\label{eq:interaction2}
\begin{split}
 S_I
 :&=\frac{\lambda}{4!}\int {\rm d}^Dx \; \varphi(x) \star \big[ \varphi(x) \star \big(\varphi(x) \star \varphi(x) \big)\big],
 \end{split}
 \end{align}
 the small-noncommutativity expansion can be readily computed to be
 \begin{align}
 \begin{split}\label{eq:interaction3}
 S_I&= \frac{\lambda}{4!} \int {\rm d}^Dx \,\left[\varphi^4 + \beta^2\, \varphi^4_{(1)}+\mathcal{O}(\beta^4)\right],
\end{split}
\end{align}
where we have defined the noncommutative correction
\begin{align}\label{eq:phi_fourth}
 \varphi^4_{(1)}:&= \frac{2}{3} \varphi^3    \Big( (D+2) +2 x^{j} \partial_{j} \Big) \partial^2\varphi.
\end{align}
Of course one could analyze all the possible different gatherings; fortunately, at this order in $\beta$ they all  turn out to be equivalent.

\subsection{Quantization and 1-loop results}
In order to perform the quantization we appeal to the functional integral formalism,
writing the action as the sum of the expansions obtained in Eqs.~\eqref{eq:kinetic} and~\eqref{eq:interaction3},
\begin{align}
 S:=S_K+S_I.
\end{align}
This allows us to define the mean field
\begin{align}
\phi(x)&:=\frac{\int \mathcal{D}\varphi\, e^{-S[\varphi]+\int dx\, J(x) \varphi(x)} \varphi(x)}{\int \mathcal{D}\varphi\, e^{-S[\varphi]+\int dx\, J(x) \varphi(x)}},
\end{align}
from which the effective action $\Gamma$ of the theory can be computed.
At the one-loop level, it can be cast as
\begin{align}\label{eq:EA_1loop}
\Gamma_{1-\text{loop}}[\phi] = S[\phi]+\frac{1}{2} \operatorname{Log} \operatorname{Det} \left(\delta^2 S[\phi]\right),
\end{align}
where the operator $\delta^2 S$ acting on an arbitrary function $f$ is defined in terms of its kernel as
\begin{align}\label{eq:d2S_operator}
\delta^2 S[\phi] f(x)&:= \int {\rm d}^D y \frac{\delta^2 S}{\delta \varphi(x)\delta \varphi(y)}[\phi] f(y).
\end{align}
An explicit computation shows that, at quadratic order in the small-deformation expansion,
the variation of the different contributions to the action can be written compactly as
\begin{align}\begin{split}\label{eq.A_I}
 \delta^2 S_I f(x) &=-\frac{\lambda}{4!}\frac{1}{2}\int  \frac{ {\rm d}^Dq_1}{(2\pi)^D}\frac{{\rm d}^Dq_2}{(2\pi)^D} \tilde{\phi}_{1} \tilde{\phi}_{2} e^{ \mathi x ( q_1+q_2)}\\
 &\times \left[4!+\beta^2\left(a_{ij}(x) (-i\partial^{i}) (-\mathi\partial^{j})+  b_{j}(x) (-\mathi\partial^{j})+c(x) \right)+\mathcal{O}(\beta^4)\right] f(x),
 \\
 \delta^2 S_K f(x) &= \Bigg(-\partial^2+ \omega^2 x^2
 -\alpha_{\rm eff}^2 x_j \partial^j \partial^i x_i+ m^2_{\text{eff}} +\mathcal{O}(\alpha^2\beta^2)\Bigg) f(x),
 \end{split}
 \end{align}
in terms of the following tensorial coefficients $a_{ij}$, $b_j$ and $c$:
 \begin{align}\label{eq.coefficients_latin}
\begin{split}
a_{ij}(x):&=8\mathi \left(2x_{i} (q_1+q_2)_{j}+ (q_1+q_2)\cdot x \delta_{ij}\right),\\
b_{j}(x):&= 8\mathi \left( x_{i} (q_1+q_2)^2+2(q_1+q_2)\cdot x\, (q_1+q_2)_{j}\right)+ 8 (2 + D)  (q_1 + q_2)_{j},\\
c(x):&=8\mathi \left( (q_1\cdot x)(2q_1\cdot q_2+q_2^2)+(q_2\cdot x)(2q_1\cdot q_2+q_1^2) \right)+ 8 (2 + D)   q_1\cdot   q_2.
\end{split}
\end{align}
Notice also that the mean field $\phi$ appears Fourier transformed in Eq.~\eqref{eq.A_I};
in effect, we use the abbreviation $\tilde \phi_i:= \tilde \phi(q_i)$ and define
 \begin{align}
\phi(x)=:\int \frac{{\rm d}^Dq_i}{(2\pi)^D} \,e^{ \mathi x q_i}\, \tilde{\phi}(q_{i}).
 \end{align}
A further remark is in order. Thanks to the expansion in the deformation parameters,
the operator $\delta^2S$ corresponds to a differential operator (in this case of second order).
Instead, the star product contains derivatives of arbitrary high order,
which means that, to tackle the problem exactly in $\alpha$ and $\beta$,
$\delta^2S$ would need to be written in terms of a pseudodifferential operator.
On physical grounds, as long as the deformation parameters remain small in comparison with the energy scale of observation
(i.e. in the renormalization flow, see below),
we can safely work within this expansion.

\subsection{Renormalization flow}

In the following we will be interested in analyzing the renormalization properties of the above-described model.
Among the various techniques available in the literature to carry out this type of computations,
in Refs.~\cite{Franchino-Vinas:2019nqy, Franchino-Vinas:2021bcl} we have employed the Worldline Formalism,
which is known for yielding compact expressions even for challenging calculations.
Recently, it has been successfully generalized to a certain type of
abelian gauge backgrounds~\cite{Copinger:2023ctz} and boundary conditions~\cite{Ahmadiniaz:2022bwy}.

In brief, the technique works as follows: first, one expresses the functional determinant in terms of a trace by means of Frullani's integral,
\begin{align}
\operatorname{Log} \operatorname{Det} \left(\delta^2 S\right)= -\int_{0}^{\infty}\frac{{\rm d}T }{T}\text{Tr} \left( e^{-T \,\delta^2 S}\right);
\end{align}
afterwards, one considers the path integral of an auxiliary first-quantized particle, which is used to compute the corresponding trace.
Instead of showing the details of the computation,
which the interested reader can consult in Refs.~\cite{Franchino-Vinas:2019nqy, Franchino-Vinas:2021bcl},
let us just focus on the divergent contributions to the effective action
in dimensional regularization.
In the two-dimensional case, the divergent pieces appear at the level of the two-point and four-point functions:
\begin{align}
\begin{split}
\Gamma^{(2)}_{D=2,\,\text{div}}&=\frac{\lambda}{24\pi (2-D) }\int {\rm d}^2x \left [
\frac{6 \alpha ^2 m^2}{\omega ^2}+4 \beta ^2 m^2+3+x^2 \left(9 \alpha ^2+8 \beta ^2 \omega ^2\right)
\right] \phi^2(x),
\\
\Gamma^{(4)}_{D=2,\,\text{div}}&=
\frac{\lambda^2}{48\pi(2-D)}\int {\rm d}^2x \left[ \frac{3 \alpha ^2}{\omega ^2}+2 \beta ^2 \right] \phi^4(x).
\end{split}
\end{align}
This implies that the counterterms which are necessary to absorb the infinities are already available in the classical action.
On the other hand, in the four-dimensional case we need to introduce further counterterms,
in order to accommodate all the following divergences in the two- and four-point functions:
\begin{align}\label{eq.2point_div}
\begin{split}
\Gamma^{(2)}_{D=4,\,\text{div}}&=-\frac{\lambda}{192\pi^2(4-D) }\int {\rm d}^4x \,\phi \,\Bigg[-52\alpha_{\text{eff}}^2 + 6 m^2 + 12 m^4 \beta^2 + \frac{6 \alpha_{\text{eff}}^2 m^4}{\omega^2} + 48 \beta^2 \omega^2
\\
&\hspace{-1.3cm}+ x^2 (15 \alpha_{\text{eff}}^2 m^2 + 6 \omega^2 + 36 m^2 \beta^2 \omega^2 - 4 \beta^2 \omega^2 \partial^2) - 8 \beta^2 \omega^2  x_{j} x_{i} \partial{}^{j} \partial^{i}
+ x^4 (9 \alpha_{\text{eff}}^2 \omega^2 + 24 \beta^2 \omega^4) \Bigg]\phi,
\end{split}
\\
\begin{split}
  \Gamma^{(4)}_{D=4,\,\text{div}}&= \frac{1}{4!}\frac{3  \lambda ^2 }{16 \pi ^2 (4-D)}
  \\
  &\hspace{-1cm}\times \int {\rm d}^4x\, \Bigg\lbrace \phi^2\, \left[ \frac{\alpha_{\text{eff}} ^2 \partial^2}{2 \omega ^2}-\frac{ 2 m^2  \left(\alpha_{\text{eff}}´ ^2+2 \beta ^2 \omega ^2\right)+\omega ^2 }{\omega ^2}-\frac{ x^2 \left(5 \alpha_{\text{eff}} ^2+16 \beta ^2 \omega ^2\right)}{2 } \right.\Bigg] \phi^2 \label{eq.4point_div}
  -\beta ^2 \phi^4_{(1)}
  \Bigg\rbrace.
  \end{split}
\end{align}
The occurrence of $x^4\phi^2$ and $x^2\phi^4$ in these expressions can be interpreted
as the need to renormalize contributions coming from the expansion of an invariant measure in curved space,
i.e. in terms as $\int \sqrt{-g} \phi^n$, $n=2,4$.
Additionally, the contribution $x_{j} x_{i} \partial{}^{j} \partial^{i}$ can be understood
as a quantum modification of the Laplace--Beltrami operator.

After introducing the necessary counterterms,
one can easily compute the beta functions associated with each of the renormalized couplings; the explicit equations can be found in Refs.~\cite{Franchino-Vinas:2019nqy, Franchino-Vinas:2021bcl}.
For instance, the beta functions of the main couplings in $D=4$ read
\begin{align}\label{eq.system}
\begin{split}
 \beta_{\lambda} &= \frac{3 \lambda^2}{16 \pi^2 } \left(1 + 4 m^2 \beta^2+\frac{2 \alpha_{\text{eff}}^2 m^2}{\omega^2}  \right),
 \\
 {\beta}_{{\omega}^2}&=\frac{\lambda}{96 \pi^2} (15 \alpha_{\text{eff}}^2 m^2 + 6 \omega^2 + 36 m^2 \beta^2 \omega^2),\\
 {\beta}_{{m}^2_{\text{eff}}} &=\frac{\lambda}{96 \pi^2} \left(-26 \alpha_{\text{eff}}^2 + 6 m^2 + 12 m^4 \beta^2 +  \frac{6 \alpha_{\text{eff}}^2 m^4}{\omega^2} + 48 \beta^2 \omega^2\right),\\
 {\beta}_{{\alpha}_{\text{eff}}^2} &=\frac{\lambda }{12 \pi^2}\beta^2 \omega^2,\\
 {\beta}_{{\beta}^2}&=-\frac{3\lambda }{8 \pi^2 }  m^2 \beta^2 \left(\frac{\alpha_{\text{eff}}^2}{\omega^2} + 2 \beta^2 \right),
 \end{split}
\end{align}
whereas the correct Callan--Symanzik equations are obtained by adding a term proportional to the couplings' classical dimensions~\cite{peshkin}.

Due to the complexity of the system of differential equations, an exact, analytic solution is not available to us.
Still, we are able to provide information on two different levels.
First, it is easy to prove that the renormalization group equation in our one-loop approximation lacks fixed points other than the trivial Gaussian one.
In this sense, the model is thus not able to provide a physical explanation to the harmonic term that, added by hand,
is responsible for the several advantages shown by the Grosse--Wulkenhaar model, for which a line of fixed points exists~\cite{Grosse:2019qps,Grosse:2005da}.

Second, as a way to gain understanding on the behaviour of the QFT,
we can solve the system numerically,
for which we will use the function \emph{odeint} of the \emph{scipy} package~\cite{scipy}.
Of course this entails providing the initial conditions at a given scale.
For the sake of brevity, let us define the vector of parameters $v=\left(\lambda,\,m_{\rm eff},\,\alpha^2,\,\beta^2,\,\omega^2 \right)$;
from now on all the quantities will be measured in units of eV and the initial condition will be imposed at the scale $\mu_0=1$.

If both $\alpha^2$ and $\beta^2$ are positive, there are no surprises,
the behaviour being analogous to the one observed in the commutative quartic theory.
In fact, this can be expected from the beta function of the self-interacting coupling $\lambda$.
More explicitly, from Eq.~\eqref{eq.system} we see that, in $D=4$ and for small couplings, $\lambda$ will grow slightly faster than in the commutative case;
whether in the full model this behaviour corresponds to a trivial theory or a Landau pole remains to be seen.

\begin{figure}
\begin{center}
\begin{minipage}{0.49\textwidth}
 \includegraphics[width=1.0\textwidth]{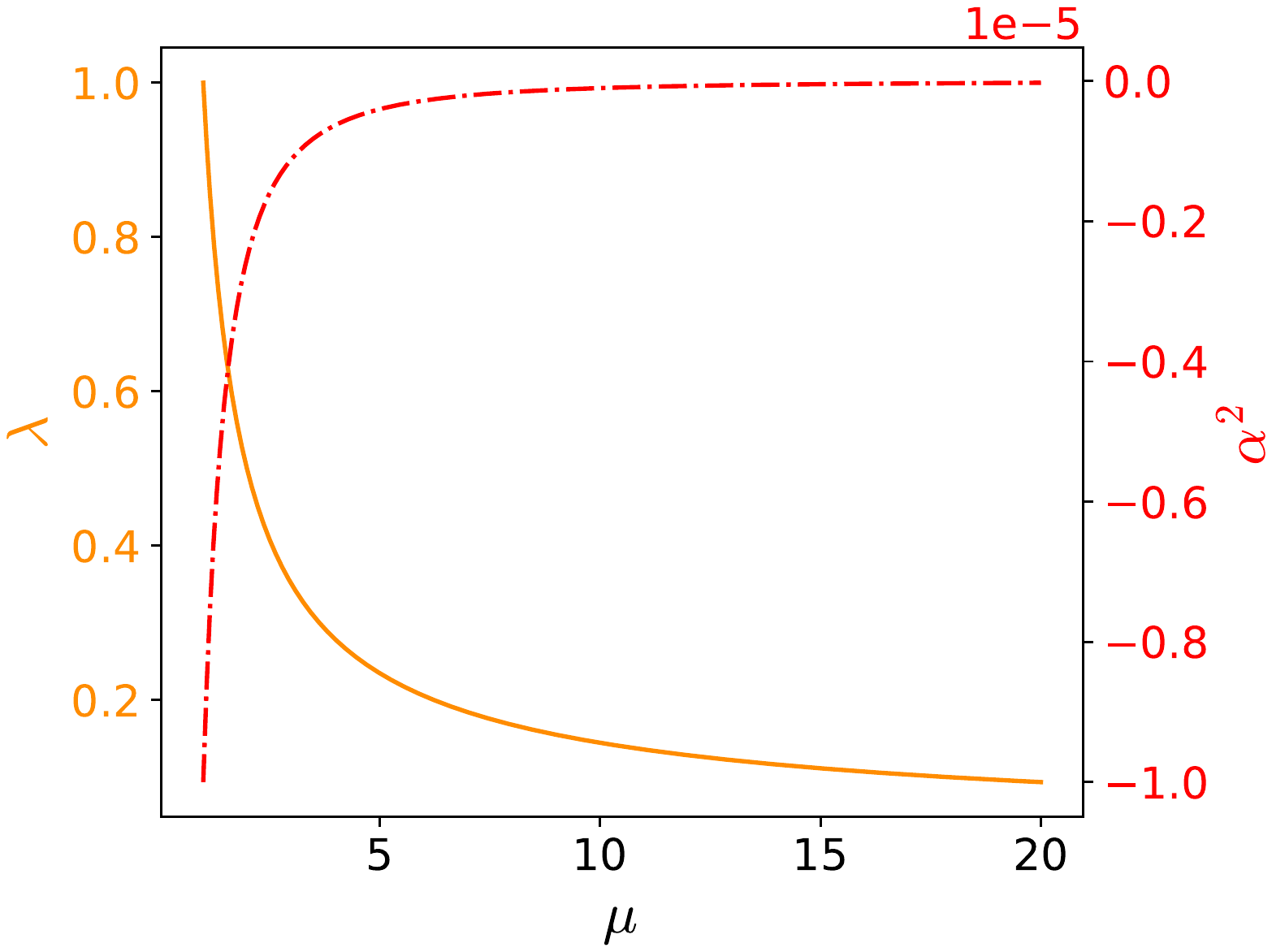}
 \end{minipage}
 \begin{minipage}{0.49\textwidth}
 \includegraphics[width=0.9\textwidth]{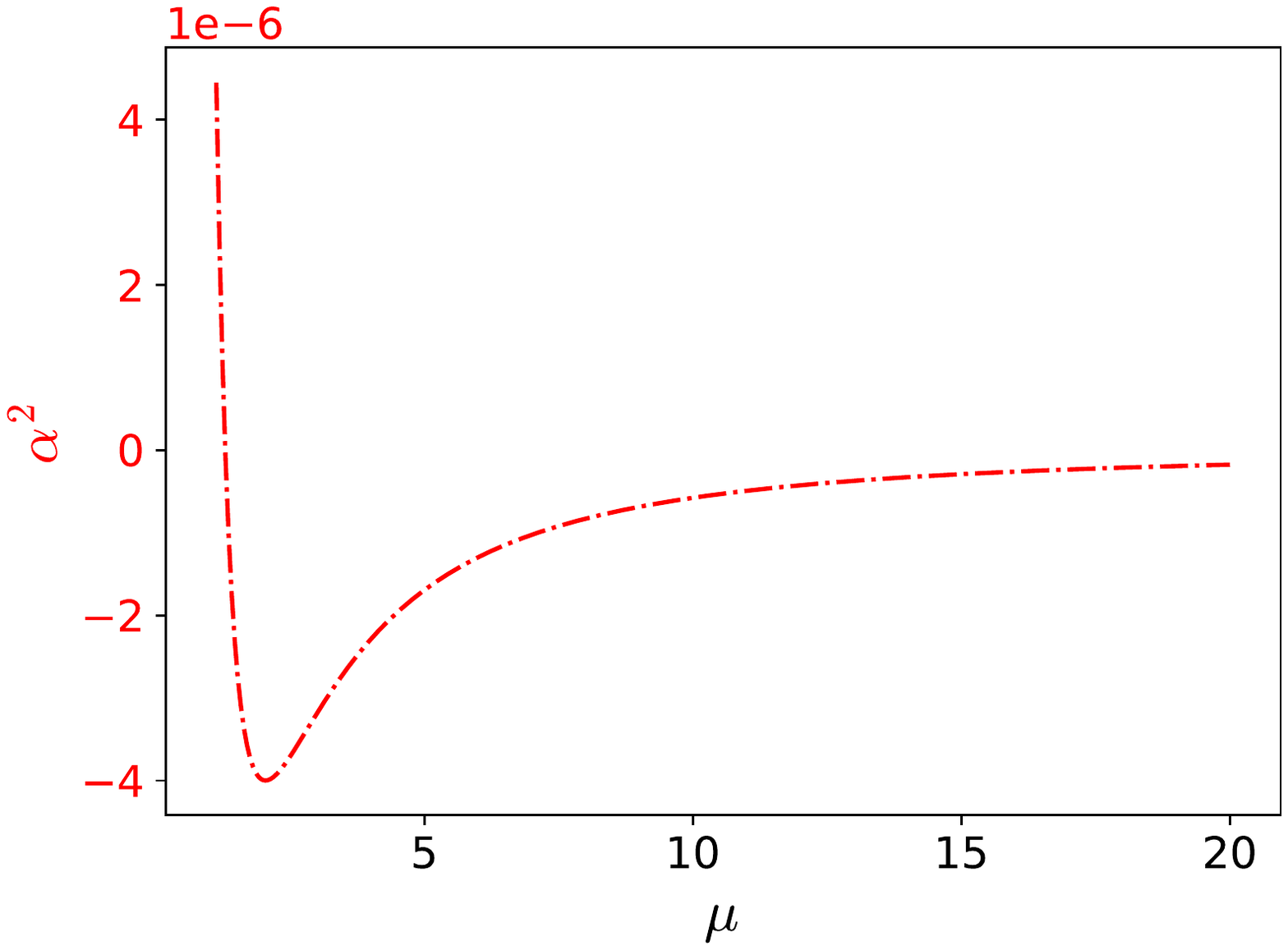}
 \end{minipage}
 \caption{Numerical solutions for the running of the self-coupling $\lambda$ (continuous, orange curve) and the parameter $\alpha^2$ (dotted-dashed, red curve) in the four-dimensional model. The initial conditions correspond to
  $v_1=\left(1,\,10,\,-10^{-5},\,- 10^{-1},\, 10^{-1}\right)$ (left panel) and $v_2=\left(1,\,10,\,10^{-5},\,- 10^{-1},\, 10^{-1}\right)$ (right panel).}
 \label{fig.functions4}
\end{center}
\end{figure}

Instead, if we allow for changes in the sign of $\alpha^2$ or $\beta^2$,
the scenario becomes much richer, thanks to the alternating signs of the single contributions to the beta functions.
For instance, consider $v_1=\left(1,\,10,\,-10^{-5},\,- 10^{-1},\, 10^{-1}\right)$ in $D=4$,
which means that we are actually considering anti-Snyder space in conjunction with an effective anti-de Sitter curved space.
In this configuration, the noncommutative contributions are large enough to overcome the positive contributions to the beta function of the self-interaction coupling;
the result is a considerable modification in the behaviour of the latter, rendering the theory asymptotically free in this one-loop analysis (see the orange, continuous curve in the left panel of Fig.~\ref{fig.functions4}).
At the same time, the effective mass, $\alpha^2_{}$ and $\omega$ tend to zero for large scales, while $\beta^2$ is the only relevant parameter of the theory in the UV (ultraviolet) regime;
as expected, the range of validity of our calculations will thus be determined by $\beta^2$.
The plot of $\alpha^2_{}$ is also offered as an example in the left panel of Fig.~\ref{fig.functions4} (red, dotted-dashed curve).

Another intriguing situation arises, for example, for $v_2=\left(1,\,10,\,10^{-5},\,- 10^{-1},\, 10^{-1}\right)$ in $D=4$,
which could be associated with a Snyder--anti-de Sitter space.
Around the starting point, the classical scaling of $\alpha^2$ gives a contribution stronger than that of the quantum corrections,
triggering a change in the sign of $\alpha^2_{}$ around the starting scale, as observed in the right panel of Fig.~\ref{fig.functions4}.
Far from $\mu=1$, $\alpha^2$ increases in absolute value, leaving the regime in which our approximations are expected to be valid.
More specifically, it becomes relevant in the IR (infrared) regime, while for increasing energy scales it develops a global minimum before tending to zero in the UV.
This phenomenon is of cosmological interest, since a sign switch of the cosmological constant has been proven
to alleviate the Hubble constant tension under certain assumptions~\cite{Akarsu:2019hmw, Akarsu:2021fol}.

Although in the two-dimensional case the observed effects are of different nature, we can appreciate the imprints of noncommutativity.
First of all, note that the commutative, flat case possesses a Gaussian fixed point which is an attractor in the UV.
Once we switch on the noncommutativity, whether $\beta^2$ is relevant or irrelevant depends on the initial conditions (also in the curved case).
The interested reader is forwarded to Ref.~\cite{Franchino-Vinas:2021bcl} for a complete explanation.



\section{Outlook}

The Snyder--de Sitter space offers several desirable features that a quantum gravity description of our universe
is expected to exhibit.
The analysis carried out for a self-interacting, quantum scalar field theory has revealed
some novelties with respect to the commutative case, leaving some open questions.

First, at the one-loop level we have shown that a harmonic term is automatically generated,
which at first sight seems to imply a connection with the Grosse--Wulkenhaar model.
However, the renormalization properties of our model seem to be more intrincated and there are no nontrivial fixed points.
One intriguing question is whether one could perform a nonperturbative analysis in SdS;
for example, one could use the analytic functional renormalization group or numerical techniques,
in a way analogous to recent studies of the Grosse--Wulkenhaar setup~\cite{Prekrat:2022sir, Prekrat:2023fts}.

Recently, generalizations of the SdS spacetime that also
include the related Yang model and its $\kappa$-deformations
have been discussed~\cite{Meljanac:2022qhp, Meljanac:2022vjb, Lukierski:2023gxf, Martinic-Bilac:2024wew}.
It will be interesting to extend our discussion to these setups;
of course, one should first construct the theory of free quantum fields on them.

Third, the nonassociative character of the involved star product plays no role in the
small-deformation, one-loop computation performed in this manuscript.
The expectation is that nonassociativity should affect our results in an all-order calculation,
so that the knowledge gathered during recent years in the understanding of nonassociative structures
would be needed~\cite{Schupp:2023jda, Aschieri:2017sug, Kupriyanov:2015dda}.

Finally, the introduction of fields with other spins is still a pending matter.
As a first approach, it could prove useful to make a step backwards and consider the introduction of spinning particles on these spaces,
for which one could use the joint machinery of the worldline formalism~\cite{Fecit:2024jcv} and curved momentum space~\cite{Franchino-Vinas:2020umq, Franchino-Vinas:2022fkh,Franchino-Vinas:2023ojp}.


 \acknowledgments
The authors are grateful to F.~Lizzi, D.~Prekrat, E.~Skvortsov and H.~Steinacker for their questions and comments.
 SAF acknowledges the support of Helmholtz-Zentrum Dresden-Rossendorf, Project 11/X748 (UNLP)
 and
 PIP 11220200101426CO (CONICET).
The authors would like to acknowledge the contribution of the COST Action CA18108 ``Quantum gravity phenomenology in the multi-messenger approach''.


 \appendix

\bibliographystyle{JHEP}
\bibliography{bibliografia}

\end{document}